\documentclass[prl,twocolumn,showpacs,floatfix,amsfonts]{revtex4}
\usepackage{graphicx,graphics,color,epsfig}
\usepackage{bm}
\usepackage{amsmath}
\usepackage{amssymb}
\usepackage{fancyheadings}
\newcommand{\beqa}{\begin{eqnarray}}
\newcommand{\eeqa}{\end{eqnarray}}

\begin{document}
\preprint{}
\title{Novel Spin Dynamics in a Josephson Junction}
\author{Jian-Xin Zhu,$^{1}$ Zohar Nussinov,$^{1}$ Alexander Shnirman,$^{1,2}$
Alexander V. Balatsky$^{1}$}
\affiliation{$^{1}$ Theoretical Division, Los Alamos National Laboratory,
Los Alamos, New Mexico 87545}
\affiliation{$^{2}$
Institut f\"ur Theoretische Festk\"orperphysik,
Universit\"at Karlsruhe,
D-76128 Karlsruhe, Germany}
\begin{abstract}
We address the dynamics of a single spin embedded in the tunneling
barrier between two superconductors. As a consequence of pair
correlations in the superconducting state, the spin displays a
rich and unusual dynamics. To properly describe the time evolution
of the spin we find the generalized Wess-Zumino-Witten-Novikov
term in the effective action for the spin on the Keldysh contour.
The superconducting correlations lead to an effective spin action
which is non-local in time leading to unconventional precessions.
Our predictions might be directly tested for macroscopic spin
clusters.
\end{abstract}
\maketitle

There is a growing interest in a number of techniques that allow
one to detect and manipulate a single spin in the solid state. A
partial list includes optical detection of electron spin resonance
(ESR) in a single molecule~\cite{Koehler93}, tunneling through a
quantum dot~\cite{Engel01}, and, more recently, ESR-scanning
tunneling microscopy (ESR-STM) technique~\cite{Mana89,Durkan02}.
Interest in ESR-STM lies in the possibility to detect and
manipulate a single spin~\cite{Manoharan02,Balatsky02}, which is
crucial in spintronics and quantum information processing. In a
similar investigation much work has been done to address the
coupling, feedback effects, and decoherence in a coupled
electronic-vibrational systems, such as nanomechanical oscillators
and  local vibrational modes~\cite{Mozy02b}. In a previous
publication, two of us studied the effect of a precessing spin on
the supercurrent~\cite{zb}. In the present Letter, we complement
this earlier study and focus on how a spin is affected by a
supercurrent. Our major finding is that the spin dynamics is no
longer that of simple precession. We provide a direct prediction
of the expected spin dynamics. This unusual spin dynamics
characterized by longitudinal oscillations is caused by coupling
to a Josephson current. Due to spin normalization, these novel
spin precessions do not lead to any changes in the Josephson
current. Keldysh contour calculations illustrate that a non-local
in time single fermion action is also found in situations wherein
the single spin cluster is replaced by an Anderson
impurity~\cite{avishai}. As well known, in the limit of small
hopping amplitudes to and from an Anderson impurity, the impurity
attains a Kondo like character much like that of the single spin
which is the focus of our attention. Here we consider  the origin
of this rather generic non-locality in time present in the
dynamics of a Josephson junction. Our primary focus will be on
larger spin clusters for which direct measurements can be made to
probe this novel spin dynamics.

\begin{figure}
\centerline{\psfig{figure=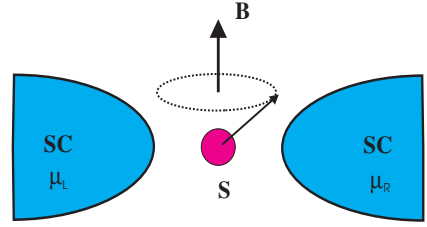,height=3.0cm,width=5.5cm,angle=0}}
\caption{Magnetic spin coupled to two superconducting leads. In
the presence of a magnetic field ${\bf B}$, the spin precesses
around the field direction. As we illustrate
here, in a Josephson junction
the standard precession is accompanied
by a polar motion leading to a full blown
rigid body like dynamics.} \label{FIG:SETUP}
\end{figure}
The model system under consideration is illustrated in
Fig.~\ref{FIG:SETUP}. It consists of two ideal superconducting
leads coupled to each other by a single magnetic spin. In the
presence of a magnetic field, the spin precesses around the field
direction.  We  neglect the interaction of the spin with two
superconducting leads.  The Hamiltonian for the Josephson junction
can be written as~\cite{Ambegaokar63,Sigrist91}:
$H=H_{L}+H_{R}+H_{T}\;.$
The first two terms are the Hamiltonians for electrons
in the left and right superconducting leads
$H_{L(R)}=\sum_{k(p);\sigma}\epsilon_{k(p)}
c_{k(p),\sigma}^{\dagger}c_{k(p),\sigma}
 +\frac{1}{2}\sum_{k(p);\sigma,\sigma^{\prime}}
[\Delta_{\sigma\sigma^{\prime}}(k(p)) c_{k(p),\sigma}^{\dagger}
c_{-k(-p),\sigma^{\prime}}^{\dagger} +\mbox{H.c.}]\;,$ where we
have denoted the electron creation (annihilation) operators in the
left (L) lead by $c_{k\sigma}^{\dagger}$ ($c_{k\sigma}$) while
those in the right (R) lead by $c_{p\sigma}^{\dagger}$
($c_{p\sigma}$). The quantities $k$ ($p$) are momenta, $\sigma$
the spin index, and $\epsilon_{k(p),\sigma}$,
$\Delta_{\sigma\sigma^{\prime}}(k(p))$ are, respectively, the
single particle energies of conduction electrons, and the pair
potential in the leads. In this work, we consider $s$-wave pairing
symmetry in the superconducting leads. The two leads are weakly
coupled via the tunneling Hamiltonian:
$H_{T}=\sum_{k,p;\sigma,\sigma^{\prime}}
[T_{\sigma\sigma^{\prime}}(k,p)
c_{k\sigma}^{\dagger}c_{p\sigma^{\prime}} +\mbox{H.c.}]\;,$
where the matrix element $T_{\sigma\sigma^{\prime}}(k,p)$ transfer
electrons through an insulating barrier. When a spin is embedded
in the tunneling barrier, the tunneling matrix becomes a spin
operator~\cite{Balatsky02}:
$\hat{T}=T_0 \delta_{\sigma\sigma'} + T_1 \mathbf{S} \cdot
\bm{\sigma}_{\sigma \sigma'}\;,$
where $T_0$ is a spin-independent tunneling matrix element and
$T_1$ is a spin-dependent matrix element originating from the
direct exchange coupling $J$ of the conduction electron to the
localized spin ${\bf S}$. We take both to be momentum independent.
This is not a crucial assumption and is merely introduced to
simplify notations. Typically, from the expansion of the work
function for tunneling, $\frac{T_1}{T_0} \sim J/U$, where $U$ is a
spin-independent tunneling barrier~\cite{zb}. We further allow a
weak external magnetic field $B \sim 10^2\; \mbox{Gauss}$. Such a
small field will not influence the superconducting state and we
may ignore its effect on the leads. The Josephson junction with
the
 spin has two time scales: (i) The Larmor
 precession frequency of the spin
 $\omega_L = g \mu_B B$, where $g, \mu_B$ are the gyromagnetic ratio and
 Bohr magneton of the conduction electron, respectively. (ii) The
frequency $\omega_{J} = 2 eV$  characterizing the
Josephson effect when an external
  voltage $V$ is applied.



We now derive the effective action via the Keldysh technique. If
all external fields are the same on both forward and backward
branches of the Keldysh contour ($C$) then $\mathcal{Z} =
\mbox{Tr}\, T_C\,\exp [-i \oint_C dt H_{T}(t)]  = 1$, where the
trace is over both the electron and the spin degrees of freedom.
We first take a partial trace in
$\mathcal{Z}$ over the lead fermions (the bath)
to obtain an effective spin action.
The Josephson contribution to the resulting spin
action reads $i\delta\mathcal{S} = - \frac{1}{2} \oint_C dt \oint_C dt'
\langle T_C H_{T}({\mathbf S(t)},t) H_{T}({\mathbf
S(t^{\prime})},t^{\prime}) \rangle$, much in the spirit of
Refs.~\cite{AES_LO}.

For brevity, we set $A_{\sigma,\sigma^{\prime}} \equiv \sum_{k,p}
c_{k\sigma}^{\dagger} c_{p\sigma^{\prime}}$. The tunneling
Hamiltonian of a phase (voltage) biased junction
 \beqa
 H_T &=& [T_0 \delta_{\sigma \sigma'} + T_1 \mathbf{S} \cdot
 \bm{\sigma}_{\sigma \sigma'}]  \nonumber \\
 &&\times \big(A_{\sigma
\sigma^{\prime}} \exp(i\phi/2) + A_{\sigma \sigma^{\prime}}^\dag
\exp (-i \phi/2)\big)\;. \label{EQ: A1}
 \eeqa
In the presence of a dc voltage bias, $\phi = 2 eVt$. As $\phi$ is
treated classically (i.e. $\phi$ is the same on the upper and the
lower branches of the Keldysh contour), the contribution $\propto
T_0^2$ to $\delta \mathcal{S}$ vanishes. The mixed contribution
$\propto T_0 T_1$ vanishes due to the singlet spin structure of
the s-wave superconductor. The only surviving contribution reads
\beqa &i\delta\mathcal{S}  = -\frac{T_1^2}{2} \oint_C dt\oint_C
dt' \left[{\bf S}(t)\cdot \bm{\sigma}_{\alpha \beta}\right] \,
\left[{\bf S}(t') \cdot \bm{\sigma}_{\delta \gamma}\right]\times &
\nonumber \\
&\left(\langle T_C A_{\alpha \beta}(t) A_{\delta \gamma}(t')
\rangle e^{ i\frac{\phi(t) +\phi(t')}{2}} + (A,\phi\rightarrow
A^{\dag},-\phi)\right) &
 \label{EQ:Seff1}\eeqa
where we keep only the Josephson (off-diagonal) terms. The
spin structure simplifies for the s-wave case:
\beqa &&i\delta\mathcal{S}  = T_1^2 \oint_C dt\oint_C dt'
\left[{\bf S}(t)\cdot {\bf S}(t')\right] [iD(t,t')] \ ,
\label{EQ:Seff3}\eeqa where $iD(t,t')\equiv \langle T_C
A_{\uparrow\uparrow}(t) A_{\downarrow\downarrow}(t')\rangle e^{
i\frac{\phi(t) +\phi(t')}{2}} + (A,\phi\rightarrow
A^{\dag},-\phi)$. Next, we perform the standard Keldysh manipulations,
defining upper
and lower spin fields ${\bf S}^{u,l}$ residing on the
forward/backward contours and reducing the time ordered integral
over Keldysh contour to the integral over forward running time at
the cost of making the Green's function $G$ a $2\times 2$ matrix.
Finally, after a rotation to the classical and quantum components
\beqa {\bf S}_1 \equiv ({\bf S}^u + {\bf S}^l)/2, \;\; {\bf S}_2
\equiv {\bf S}^u - {\bf S}^l,\;\; {\bf S}_1 \cdot {\bf S}_2 = 0,
\label{EQ:Sfields} \eeqa we obtain $i\delta\mathcal{S} =
i\mathcal{S}_R + \mathcal{S}_I$, where
\begin{equation}
i\mathcal{S}_R=\int\int dt dt' \, {\bf S}_2(t) \cdot {\bf S}_1(t'
)[i K_{12}(t,t')]. \label{EQ:Seff2}
\end{equation}
with $K_{12}(t,t') = T_1^2\,[D^R(t,t') + D^A(t',t)]$, and
\begin{equation}
\label{Eq:S_I}
\mathcal{S}_I = \frac{1}{2} \int\int dt dt'
{\bf S}_2(t) \cdot {\bf S}_2(t')[iK_{22}(t,t')],
\end{equation}
with $K_{22}(t,t') = T_1^2\,D^K(t,t')$. The retarded (R) and
Keldysh (K) components are defined via \beqa iD^{R}(t,t^{\prime})
&= &\Theta(t-t^{\prime})\nonumber \\
&&\times \langle[A_{\uparrow\uparrow}(t),
A_{\downarrow\downarrow}(t^{\prime})]_{-}\rangle
e^{i\frac{\phi(t)+ \phi(t^{\prime})}{2}}- \mbox{c.c.}\;, \label{Eq:D_R} \\
iD^{K}(t,t') &= & \langle \{ A(t)_{\uparrow\uparrow},
A_{\downarrow\downarrow}(t') \}_{+} \rangle e^{i\frac{\phi(t) +
\phi(t^{\prime})}{2}} + \mbox{c.c.} \eeqa The advanced component
in our case is simply $iD^A(t,t') =iD^R(t',t)$. The kernels
$K_{12}$ and $K_{22}$ are readily calculated at $T=0$ from
\begin{equation}
\label{Eq:AA} \langle A(t)_{\uparrow\uparrow}
A_{\downarrow\downarrow}(t') \rangle = \sum_{k,p}
\frac{\vert\Delta\vert^{2}}{4E_{k}E_{p}} \,
e^{-i(E_{k}+E_{p})(t-t')} \; , \end{equation} where
$E_{k(p)}=\sqrt{(\epsilon_{k (p)}-E_{F})^{2}+\vert \Delta
\vert^{2}}$. As only frequencies higher than $2\Delta$ are present
in (\ref{Eq:AA}), $S_I$ (Eq.~(\ref{Eq:S_I})) vanishes for slow
fluctuations of $\mathbf{S}_2$ unlike the real part
of the action $S_R$ (due to the presence of $\Theta$ function in
Eq.~(\ref{Eq:D_R})).


\begin{figure}
\includegraphics*[width=1in]{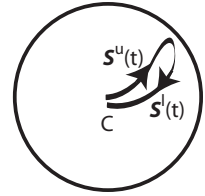}
\caption{The unit sphere for the vectors ${\bf n}^{u,l}(t)$ is
shown. The contour $C$ is the Keldysh contour for the forward ($u$)
and backward ($l$)
evolution. To properly describe the spin on this
closed contour we analyze the
WZWN term, see
Eq.~(\ref{EQ:WZWN}). For clarity, we draw a small
piece of the closed trajectories.} \label{FIG:WZWN}
\end{figure}

To properly describe the dynamics of the spin fields ${\bf S}_1$
and ${\bf S}_2$, we employ the path integral representation for
the spin fields~\cite{WZW}. The action for a free spin consists of
two terms $\mathcal{S}_{0} =  g \mu_{B} \oint_{C} dt {\bf{B}}
\cdot {\bf{S}} + \mathcal{S}_{WZWN}$. The second,
Wess-Zumino-Witten-Novikov (WZWN), term describes the Berry phase
accumulated by the spin as a result of motion of the spin on the
sphere \cite{WZW}. We generalize this action for nonequlibrium
dynamics within the Keldysh contour formalism (Fig.~\ref{FIG:WZWN}).
We write the WZWN term on both forward and backward
contours, and consequently analyze matters in terms of ${\bf S}_1$
and ${\bf S}_2$,
 \beqa i\mathcal{S}_{WZWN}=\frac{i}{S^2} \int_0^1d\tau
\int dt[{\bf S}^u(t,\tau) \cdot (\partial_{\tau}{\bf
S}^u(t,\tau)\nonumber
\\ \times \partial_t{\bf
S}^u(t,\tau)) -(u\rightarrow l)]\;. \label{WZW_diff} \eeqa The
relative minus sign stems from the backward time ordering on the
return part of $C$. The spins ${\bf S}^{u,l} = S {\bf n}^{u,l}$
with $S$ the magnitude of the spin and ${\bf n}^{u,l}$ a unit
vector field. The additional integral over $\tau$ permits us to
express the action in a local form. At $\tau =0$ we set the spin
to point along the $z$ direction at all times; at $\tau = 1$ the
spin field corresponds to the physical configurations. Each of the
individual WZWN phases (for both the forward ($u$) and backward
($l$) branches) is the spin magnitude ($S$) multiplied by the
areas spanned by the trajectories ${\bf{n}}^{u,l}(t)$ on the unit
sphere. The WZWN term contains {\em odd} powers of ${\bf S}_2$.
Insofar as the WZWN term of Eq.(\ref{WZW_diff}) is concerned, the
standard Keldysh transformation to the two classical and quantum
fields, $\mathbf{S}_{1}$ and $\mathbf{S}_{2}$, mirrors the
decomposition of the spin in an antiferromagnet (AF) to the two
orthogonal slow and fast fields \cite{explain}. The difference
between the two individual WZWN terms in Eq.(\ref{WZW_diff}) is
the area spanned between the forward and backward time
trajectories. The magnitude of this area traced between times $t$
and $t+\delta t$  is
\begin{eqnarray}
\delta \mathcal{S}_{WZWN} = S (\delta t) \delta {\bf n} \cdot ({\bf n}
\times
\partial_{t} {\bf n}).
\label{area_var}
\end{eqnarray}
Here, the variation between the forward and backward trajectory at
a given instant of time is $\delta  {\bf n} = {\bf S}_{2}/S$. In
Eq.(\ref{area_var}), we note that for small variations between the
forward and backward trajectories, ${\bf n} = {\bf S}_{1}/S$. The
WZWN action on the Keldysh loop may be expressed as
\begin{eqnarray}
\mathcal{S}_{WZWN} = \frac{1}{S^2} \int dt\, {\bf{S}}_{2} \cdot
({\bf{S}}_{1} \times \partial_{t} {\bf{S}}_{1}).
\end{eqnarray}
The real part of the total action, $\mathcal{S}_{cl}\equiv \mathcal{S}_{0} +
\mathcal{S}_{R}$, determines the quasi-classical
equation of motion. The imaginary part of the action
$\mathcal{S}_{I}$ is usually responsible for the Langevin
stochastic term. In our case, however, the Langevin term is
suppressed at frequencies much lower than $\Delta$.
The fluctuations of $\mathbf{S}_2$ are not suppressed by the
bath and no dissipation appears. The only reason to consider small
fluctuations of $\mathbf{S}_2$ is the dominance of the
quasi-classical trajectories determined by $\mathcal{S}_{cl}$.
Thus our analysis applies for large spins. The classical action
\begin{eqnarray}
\mathcal{S}_{cl} = \mathcal{S}_{WZWN}  + g \mu_{B} \int dt
{\bf{B}} \cdot {\bf{S}}_{2} \nonumber
\\
+ \int dt \int dt^{\prime} K_{12}(t,t^{\prime}) {\bf{S}}_{2} (t)
\cdot {\bf{S}}_{1}(t^{\prime}). \label{EQ:WZWN}
\end{eqnarray}
As the spin dynamics is much slower as compared to electronic
processes, we set ${\bf{S}}_1(t^{\prime}) \simeq {\bf{S}}_1(t) +
(t^{\prime} - t) d{\bf{S}}_1/dt$. The variational equations
$\delta S_{cl}/ \delta{\bf S_2}(t)= 0$ imply
\begin{eqnarray}
\frac{d {\bf{n}}}{dt} = \alpha {\bf{n}} \times \frac{d
{\bf{n}}}{dt}  \sin \omega_{J} t + g\mu_{B} {\bf{n}} \times
{\bf{B}}, \label{EQ:CLASSICAL}
\end{eqnarray}
where henceforth we denote ${\bf{S}}_{1}$ by ${\bf{S}}= S {\bf n}$,
and
$\alpha=  S \sum_{k,p}\frac{\vert \Delta\vert^{2}\vert
T_{1}\vert^{2}}{E_{k}E_{p}}[ (E_{k}+E_{p}-eV)^{-2}
-(E_{k}+E_{p}+eV)^{-2}].\;$

The non-dissipative term proportional to $\alpha$ in
Eq.~(\ref{EQ:CLASSICAL}) arises from superconducting
retardations. Both ${\bf{S}} \times \frac{d {\bf{S}}}{dt}$ and $I
\sim \sin \omega_J t$ are odd in time and their product is allowed
in the equation of motion. Due to the spectral gap, dissipation is
faint at low temperatures and frequencies.


\begin{figure}
\centerline{\psfig{figure=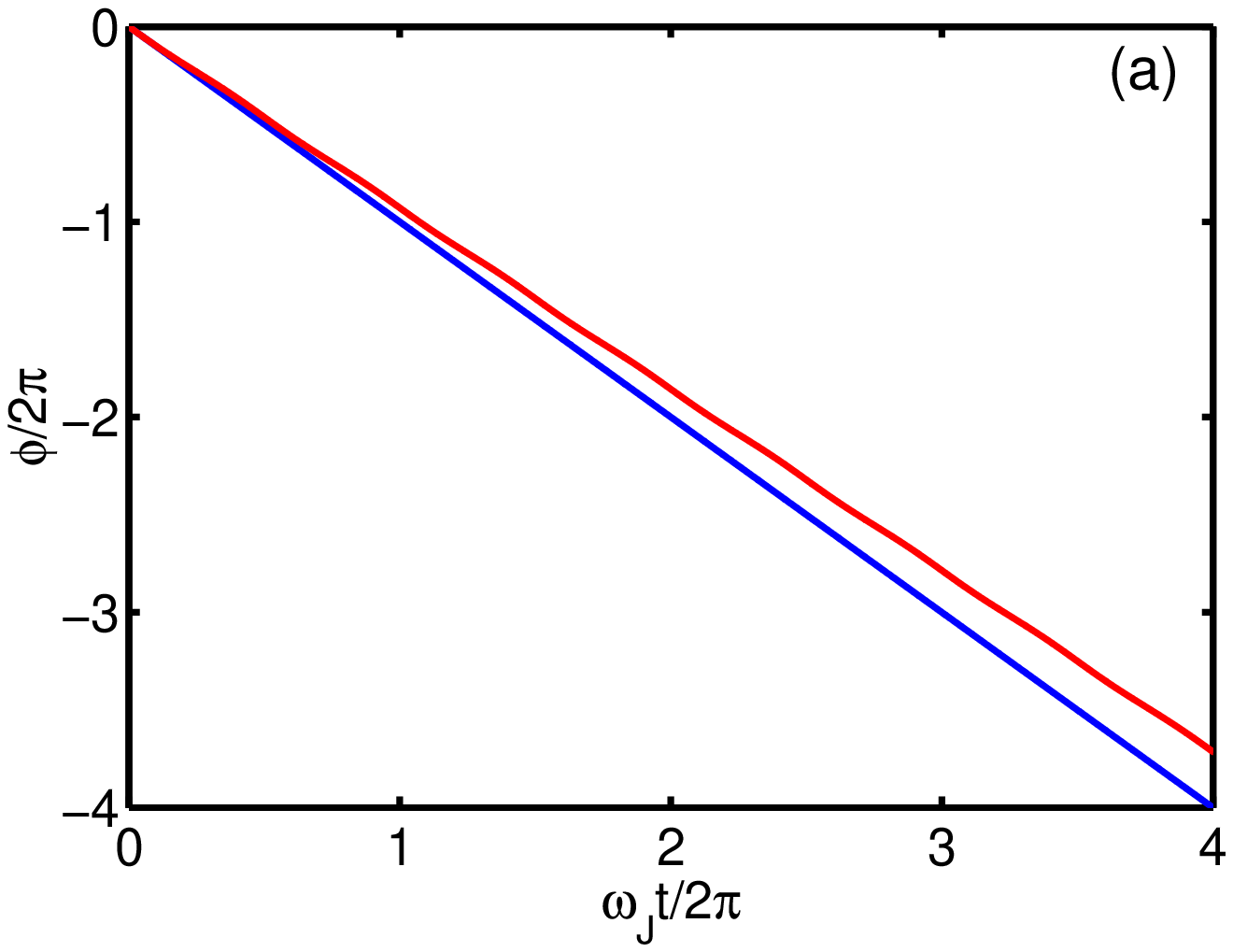,height=3.5cm,width=4.0cm,angle=0}
\psfig{figure=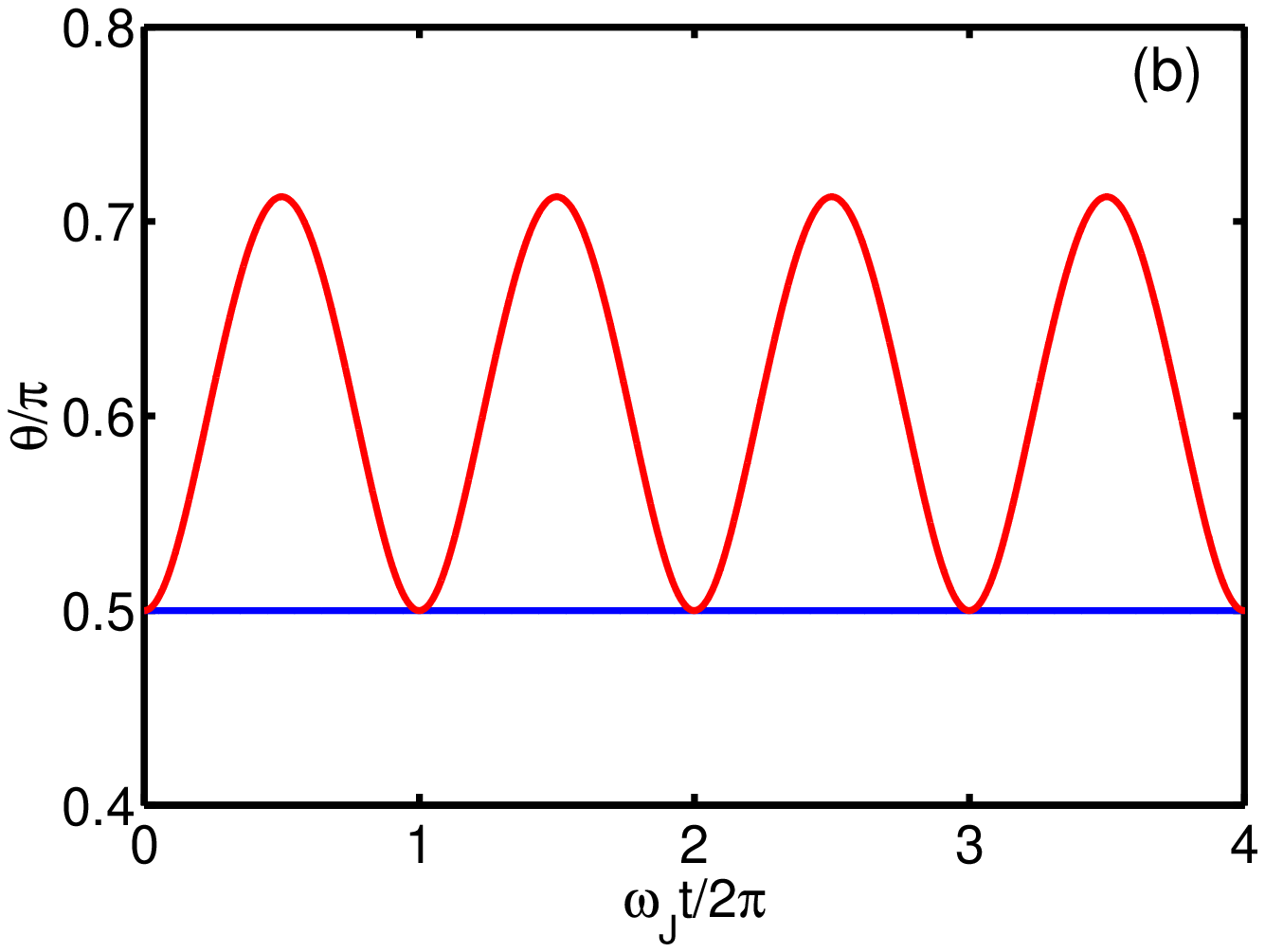,height=3.5cm,width=4.0cm,angle=0}}
\centerline{\psfig{figure=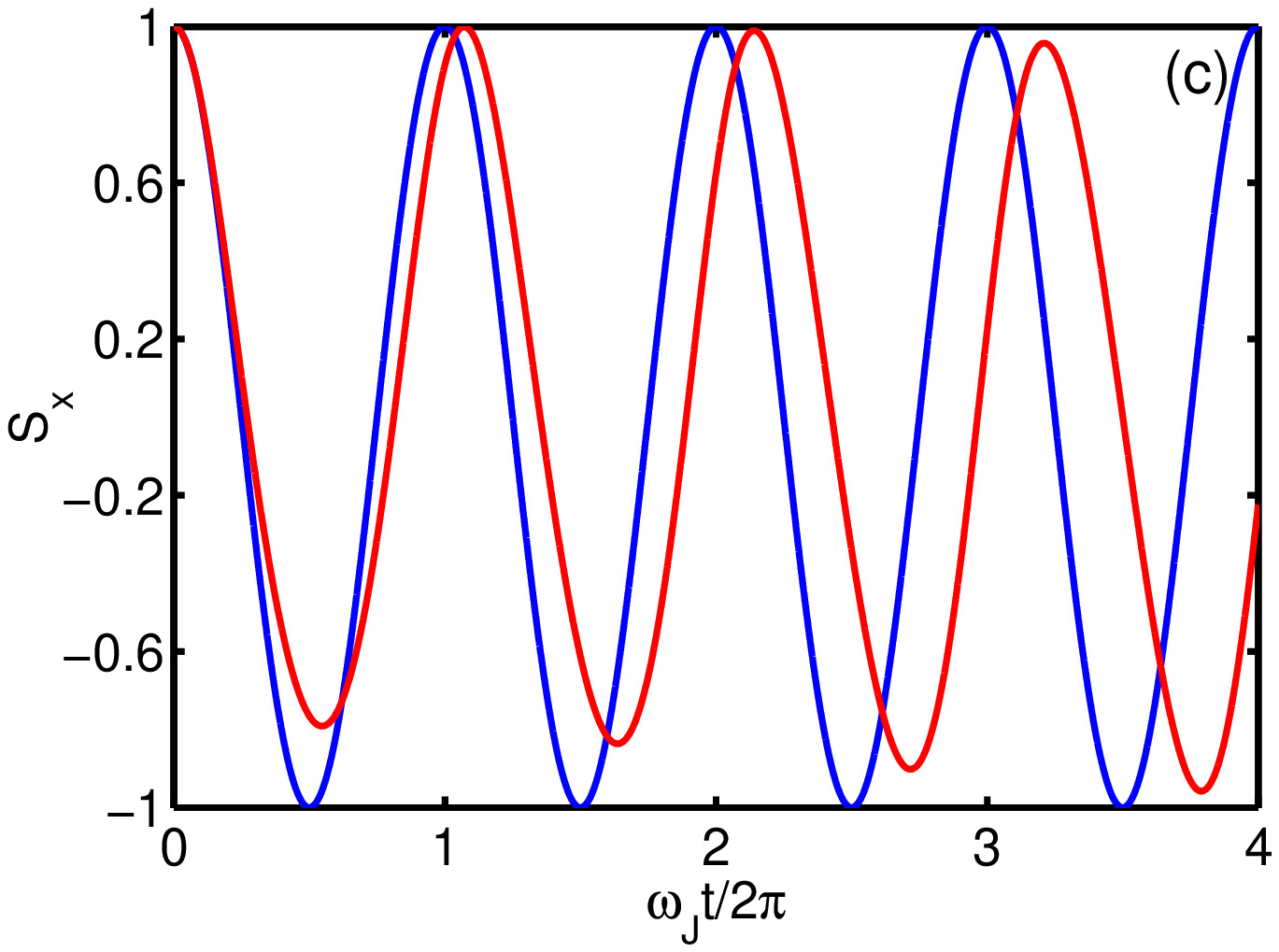,height=3.5cm,width=4.0cm,angle=0}
\psfig{figure=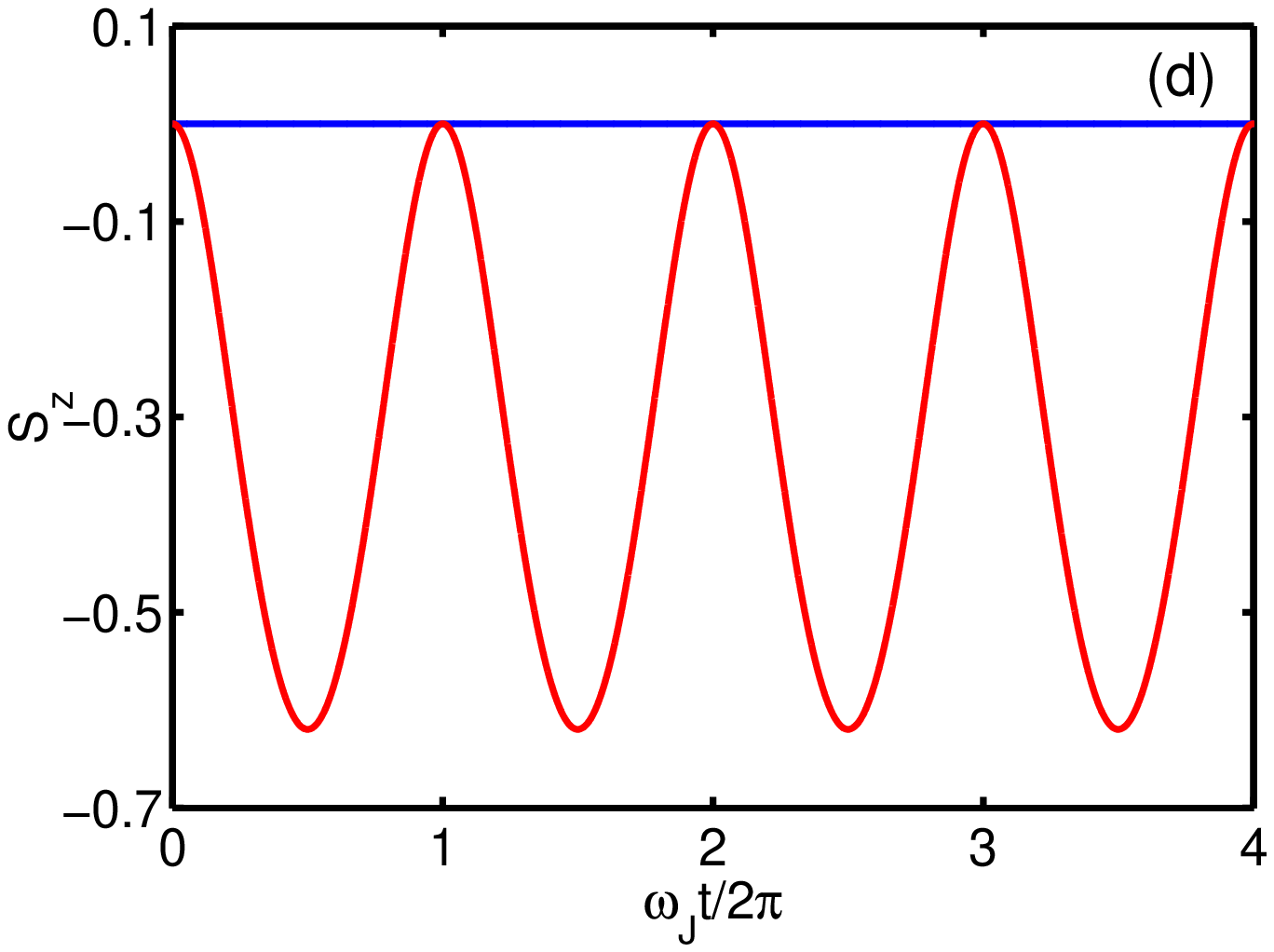,height=3.5cm,width=4.0cm,angle=0}}
\caption[*]{The spin dynamics. Top panels: the azimuthal ($\phi$)
and polar ($\theta$) angles as a function of time for an exaggerated
$\alpha = 0.4$ (red) versus the standard precessions in the
absence of the coupling ($\alpha =0$ in blue). Note that
unlike the standard case, the semi-classical spin trajectory
is not confined to planar
motions. The polar angle oscillates between its two extreme values
$\theta_{1}$ and $\theta_{2}$.
In the bottom panels we display
$\langle S_{x}(t) \rangle$ and $\langle S_{z}(t) \rangle$. Here we
take $\omega_{L}=\omega_{J}$.}
\label{FIG:TOP}
\end{figure}

\begin{figure}
\centerline{\psfig{figure=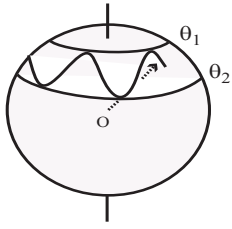,height=3.0cm,width=3.0cm,angle=0}}
\caption{The resulting spin motion on the unit
sphere in the general case.
As in the motion of classical spinning
top, the spin exhibits undulations along the polar direction.
}
\label{FIG:RIGID}
\end{figure}

The classical equation of motion, Eq.~(\ref{EQ:CLASSICAL}),
implies that ${\bf{n}} \cdot d{\bf{n}}/dt = 0$, as it must to be
consistent with the parameterization of the spin on the sphere,
$S{\bf{n}} = S(\sin \theta \cos \phi, \sin \theta \sin \phi, \cos
\theta)$. Orienting the $z$-axis along the external magnetic
field ${\bf{B}}$, Eq.~(\ref{EQ:CLASSICAL}) reads
\begin{eqnarray}
\frac{d \phi}{d t}
&=&-\frac{\omega_{L}}{1+\alpha^{2}\sin^{2}(\omega_{J}t)}\; ,
\nonumber
\\ \frac{d
\theta}{dt} &=& - \alpha \frac{d \phi}{dt} \sin \theta \sin
\omega_{J}t\; .
\end{eqnarray}
For a spin initially oriented at an angle $\theta_{0}$ relative
to ${\bf B}$,
\begin{eqnarray*}
\phi(t)  &=& - \frac{\omega_{L}}{\omega_{J}\sqrt{1+ \alpha^{2}}}
\tan^{-1}[ \sqrt{1+ \alpha^{2}} \tan(\omega_{J}t)], \nonumber
\\
\theta(t) &=& 2\tan^{-1}\Big( \biggl{[} \frac{(1-c\cos
(\omega_{J}t))(1+c)}{(1+c\cos(\omega_{J}t))(1-c)}
\biggr{]}^{\gamma} \tan \frac{\theta_{0}}{2} \Big)\;,
\end{eqnarray*}
with $c= |\alpha|/\sqrt{1+ \alpha^{2}}$ and
$\gamma=-\alpha\omega_{L}/2\omega_{J}c$. For $\alpha \ll 1$,
$\phi \simeq -\omega_{L}t$ and
$\theta \simeq \theta_0 -
\alpha (\omega_{L}/\omega_{J})\sin\theta_0\cos\omega_J t$.
Canonically, whenever a
single spin is subjected to a uniform magnetic field, the spin
precesses azimuthally with a frequency $\omega_{L}$. In a
Josephson junction, however, the spin exhibits additional polar
($\theta$) displacements. The resulting dynamics may be likened to
that of a rotating rigid top. The Josephson current leads to a
full non-planar gyroscopic motion (nutation) of the spin much like that
generated by applied ``torques'' on a mechanical top. When ${\bf
B}=0$, the solutions do not allow for any spin dynamics. The
possibility remains, nonetheless, that coupling to random magnetic
field (bath) will have a nontrivial effect on the spin in presence
of a Josephson current. In Fig.~\ref{FIG:TOP}, we display the
resulting dynamics for the spin. A schematic of the generic spin
motion is displayed in Fig.~\ref{FIG:RIGID}. Similar to a
classical spinning top, the spin wobbles along the polar direction
in addition to azimuthal rotations. Similar dynamics is
even expected for a quantum $S =1/2$.
In a spin coherent state
path integral, the
Schrodinger equation
is essentially classical.

The Josephson current in the presence of a precessing spin has
been discussed in~\cite{zb}. Within our approximation, the terms
in the action that may affect the current are proportional to
$\mathbf{S}^{2}=1$ and $\mathbf{S}\cdot \frac{\partial
\mathbf{S}}{\partial t}=0$. Consequently, the Josephson current
remains unmodulated.
Nevertheless, we may directly
monitor the spin dynamics by SQUID measurements.
The spin motion generates a time
dependent magnetic field, $ \delta {\bf B}({\bf r},t) =
\frac{\mu_{0}}{4 \pi} [3 {\bf r} ({\bf r} \cdot {\bf m}(t))
-   r^{2} {\bf m}(t)]/r^{5}$, superimposed against
the constant external field ${\bf B}$.
Here, ${\bf r}$ is the position
from the spin and the magnetic moment
${\bf m}(t) = g \mu_{B} S {\bf n}(t)$.
For a ferromagnetic cluster of spin $S= 100$, a detectable field $
\delta B \sim 10^{-10}$ Tesla appears in a SQUID placed a micron
away. For a SQUID loop of micron dimensions, the corresponding
flux variation $\delta \Phi \sim 10^{-7} \Phi_{0}$ (with
$\Phi_{0}$ a flux quantum) ---within reach of modern SQUIDs. For
such a setup with $(T_{1}/T_{0}) \sim 10^{-1}$, typical critical
Josephson current
$J_{S}^{(0)} \sim 10 \;\mu\mbox{A}$,
$\vert\Delta\vert=1.0\;\mbox{meV}$, and $eV \sim 10^{-3}
|\Delta|$, we find that $\alpha \sim 0.1$. As $n_{x}= \sin \theta
\cos \phi$, with a similar relation for $n_{y}$, the spin
components orthogonal to ${\bf B}$ vary, to first order in
$\alpha$, with a Fourier component at frequency $|\omega_{L} \pm
\omega_{J}|$, leading to a discernable signal in the magnetic
field $({\bf B} + \delta {\bf B})$.  For a field $B \sim 200\;
\mbox{Gauss}$, $\omega_{L} \sim 560\; \mbox{MHz}$ wherein for a
substantial voltage ($\omega_{J}$) range, a new side band  will
appear at $|\omega_{L} - \omega_{J}|$ whose magnitude may be tuned
to ${\cal{O}}(10-100)\;\mbox{MHz}$. This measurable frequency is
markedly different from that associated with standard Larmor
frequency ($\omega_{L}$) precessions.

{\em Conclusion.} By analyzing WZWN phases within the
non-equilibrium Keldysh framework, we find novel non-planar spin
dynamics in the presence of a tunneling Josephson current. The
coupling of the spin to a superconducting bath produces
non-damping spin retardation. This retardation results from
additional non-dissipative terms in the spin equations of motion.
There are important differences with the case of Josephson effect
in the presence of a vibrational mode \cite{Jvibration}. The
normalization of the spin disallows resonance effects at $\omega_L
= \omega_J$. The polar oscillations do not lead to additional
harmonics in the Josephson current. Nevertheless, the dynamics may
be monitored by SQUID measurements.

We thank Yu. Makhlin for useful discussions. This work was
supported by the US DOE.


\begin{thebibliography}{99}

\bibitem{Koehler93} J. Koehler {\em et al.}, Nature {\bf 363}, 342
(1993); J. Wrachtrup {\em et al.}, {\em ibid.} {\bf 363}, 244
(1993); Phys. Rev. Lett. {\bf 71}, 3565 (1993).

\bibitem{Engel01} H.-A. Engel and D. Loss, Phys. Rev. Lett. {\bf
86}, 4648 (2001); Phys. Rev. B {\bf 65}, 195321 (2002).

\bibitem{Mana89} Y. Manassen {\em et al.}, Phys. Rev. Lett.
{\bf 62}, 2531 (1989); D. Shachal and Y. Manassen, Phys. Rev. B
{\bf 46}, 4795 (1992); Y. Manassen, J. Magnetic Reson. {\bf 126},
133 (1997); Y. Manassen, I. Mukhopadhyay, and N. Ramesh Rao, Phys.
Rev. B {\bf 61}, 16223 (2000).

\bibitem{Durkan02} C. Durkan and M. E.
Welland, Appl. Phys. Lett. {\bf 80}, 458 (2002).

\bibitem{Manoharan02} H. Manoharan, Nature {\bf 416}, 24 (2002);
H. Manoharan, C. P. Lutz, and D. Eigler, Nature {\bf 403}, 512
(2000).

\bibitem{Balatsky02} A. V. Balatsky and I. Martin, Quan. Inform.
Process. {\bf 1}, 53 (2002); A. V. Balatsky, Y. Manassen, and R.
Salem, Phys. Rev B {\bf 66}, 195416 (2002).


\bibitem{Mozy02b} D. Mozyrsky and I. Martin, Phys. Rev. Lett. {\bf
89}, 018301 (2002).

\bibitem{zb} J.-X. Zhu and A. V. Balatsky, Phys. Rev. B {\bf
67}, 174505 (2003).

\bibitem{avishai} Y. Avishai, A. Golub, and A. D. Zaikin,
Phys. Rev. B {\bf 63}, 134515 (2001).



\bibitem{Ambegaokar63} V. Ambegaokar and A. Baratoff, Phys. Rev.
Lett. {\bf 10}, 486 (1963); {\bf 11}, 104(E) (1963).

\bibitem{Sigrist91} M. Sigrist and K. Ueda, Rev. Mod. Phys. {\bf
63}, 239 (1991); V. Geshkenbein and A. I. Larkin, JETP Lett. {\bf
43}, 395 (1986); D. A. Wollman  {\em et. al.}, Phys. Rev. Lett.
{\bf 71}, 2134 (1993).

\bibitem{AES_LO} U. Eckern, G. Sch\"on, and V. Ambegaokar, Phys.
Rev. B {\bf 30}, 6419 (1984); A.I. Larkin and Yu. N. Ovchinnikov,
Phys. Rev. B {\bf 28}, 6281 (1983).


\bibitem{WZW} E. Fradkin, {\em Field Theories of Condensed
Matter Systems} (Addison-Wesley, Redwood City, 1991); S. Sachdev,
{\em Quantum Phase Transitions} (Cambridge University Press,
London, 1999).

\bibitem{explain}
These two orthogonal AF fields represent (i) the slowly varying
staggered spin field (the antiferromagnetic staggered moment
$\mathbf{m}$ taking on the role of $\mathbf{S}_{1}$) and (ii) the
rapidly oscillating uniform spin field $\mathbf{l}$ (paralleling
our $\mathbf{S}_{2}$). In the antiferromagnetic correspondence,
the two forward time spin trajectories at two nearest neighbor AF
sites become the two forward ($u$) and backward ($l$) single spin
trajectories of the non-equilibrium problem. This staggered
doubling correspondence is general.



\bibitem{Jvibration} J.-X. Zhu, Z. Nussinov, and A. V. Balatsky,
cond-mat/0306107.


\end{thebibliography}
\end{document}